\documentclass[review,number,sort&compress]{elsarticle}
\usepackage{lineno}
\biboptions{sort&compress}

\journal{Journal of \LaTeX\ Templates}

\begin{document}

\begin{frontmatter}

\title{Measurement of gamma quantum interaction point in plastic scintillator with WLS strips}

\author[add1]{J. Smyrski\corref{mycorrespondingauthor}}
\cortext[mycorrespondingauthor]{Corresponding author}
\ead{smyrski@if.uj.edu.pl}
\author[add1]{D.~Alfs}
\author[add1]{T.~Bednarski}
\author[add1]{P.~Bia{\l}as}
\author[add1]{E.~Czerwi{\'n}ski}
\author[add1]{K.~Dulski}
\author[add1]{A.~Gajos}
\author[add1]{B.~G{\l}owacz}
\author[add1]{N.~Gupta-Sharma}
\author[add2]{M.~Gorgol}
\author[add2]{B.~Jasi{\'n}ska}
\author[add1]{M.~Kajetanowicz}
\author[add1]{D.~Kami{\'n}ska} 
\author[add1]{G.~Korcyl}
\author[add3]{P.~Kowalski}
\author[add4]{W.~Krzemie{\'n}}
\author[add1]{N.~Krawczyk}
\author[add1]{E.~Kubicz}
\author[add1]{M.~Mohammed}
\author[add1]{Sz.~Nied{\'z}wiecki}
\author[add1]{M.~Pawlik-Nied{\'z}wiecka}
\author[add3]{L.~Raczy{\'n}ski}
\author[add1]{Z.~Rudy}
\author[add1]{P.~Salabura}
\author[add1]{M.~Silarski}
\author[add1]{A.~Strzelecki}
\author[add1]{A.~Wieczorek}
\author[add3]{W.~Wi{\'s}licki}
\author[add1]{J.~Wojnarska}
\author[add2]{B.~Zgardzi{\'n}ska}
\author[add1]{M.~Zieli{\'n}ski}
\author[add1]{P.~Moskal}

\address[add1]{Faculty of Physics, Astronomy and Applied Computer Science, Jagiellonian University,
   S.~{\L}ojasiewicza 11, 30-348 Cracow, Poland}

\address[add2]{Department of Nuclear Methods, Institute of Physics,
  Maria Curie-Sklodowska University, 20-031 Lublin, Poland}

\address[add3]{{\'S}wierk Computing Centre, National Centre for Nuclear Research,
   05-400 Otwock-{\'S}wierk, Poland}

\address[add4] {High Energy Department, National Centre for Nuclear Research,
   05-400 Otwock-{\'S}wierk, Poland}

\begin{abstract}

We demonstrated the feasibility of measuring the axial coordinate
of a gamma quantum interaction point in a plastic scintillator bar
via the detection of scintillation photons escaping from the scintillator 
with an array of wavelength-shifting (WLS) strips.
Using a test set-up comprising a BC-420 scintillator bar 
and an~array of sixteen BC-482A WLS strips 
we achieved a spatial resolution of~5\,mm~($\sigma$)  for annihilation photons from a $^{22}$Na isotope.
The studied method can be used to improve the spatial resolution
of a plastic-scintillator-based PET scanner which is being developed by the J-PET collaboration.

\end{abstract}
\corref{mycorrespondingauthor}

\end{frontmatter}

%\linenumbers

\section{Introduction}

Positron emission tomography (PET) traditionally uses inorganic scintillation crystals for detection
of the annihilation photons~\cite{Slonka_SLM2016},
and progress in the PET technology is closely connected with the developments 
in the field of scintillation crystal detectors.
However, one can also try to improve some of the PET performance parameters such as the time resolution
or the spatial resolution by replacing the inorganic crystals with different types of detectors like 
straw tubes~\cite{Shehad_IEEE2005,Sun_IEEE2007},
resistive plate chambers (RPCs)~\cite{Blanco_IEEE2006}, silicon pad detectors~\cite{Park_NIM2007} 
or~liquid xenon detectors \cite{Doke_NIM2006, Amaudruz_NIM2009, Miceli_PMB2012}.

The J-PET collaboration developed  a whole-body PET scanner 
based on plastic scintillators which will allow
to reach a superior time-of-flight (TOF) resolution and a~high spatial acceptance at a moderate price.
The scanner consists of plastic scintillator bars read out at both ends by
a pair of photomultipliers and arranged axially around a cylindrical
tomograph tunnel~\cite{Moskal_BAMS2011, Moskal_NMR2012, Moskal_NIM2014, Moskal_NIM2015, Moskal_PMB2016}.
The axial coordinate of~the annihilation photon interaction point in the scintillator bar is derived from 
the difference of the light propagation time measured with the pair of photomultipliers.

Even with a high precision of the time difference measurement of 140\,ps~($\sigma$), obtained with a prototype
of the J-PET scanner,
the corresponding position resolution in the axial direction is only moderate and amounts 
to about 10\,mm~($\sigma$)~\cite{Sharma_NUKA2015}. 
This only improves to about 9\,mm when applying multi-threshold readout and signal reconstruction 
based on the mathematically advanced compressing sensing theory~\cite{Raczynski_NIM2014, Raczynski_NIM2015}.
Therefore, in order to obtain higher resolution, we~propose to register 
scintillation light escaping the scintillator bar through a~side wall 
using an array of wavelength-shifting (WLS) strips \cite{Smyrski_BAMS2014}.

\begin{figure}
\centering
\includegraphics[width=0.9\linewidth]{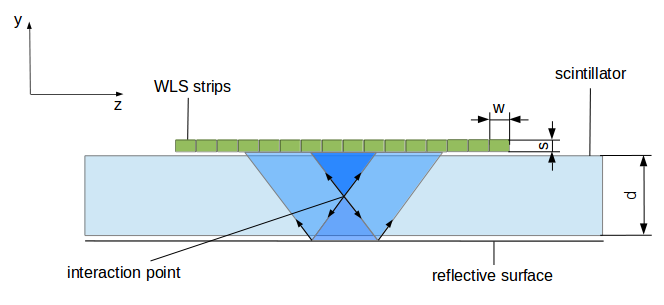}
\caption{Principle of measuring the axial coordinate
of the gamma quantum interaction point in a~plastic scintillator bar
using an array of WLS strips. 
The z-axis of indicated coordinate system points
along the scintillator bar and the y-axis is orthogonal to a side wall of the scintillator.
The WLS strips are oriented orthogonally to the y-z-plane.
} 
\label{fig:principle}
\end{figure}

A schematic view of the setup is shown in~Fig.~\ref{fig:principle}.
It shows the plastic scintillator bar 
%A principle of the proposed method of position determination is explained in~Fig.~\ref{fig:principle}. 
%It shows a schematic view of a set-up consisting of a plastic scintillator bar 
and the array of WLS strips placed near the scintillator.
%Scintillation photons emitted within a cone limited by the angle of the total internal reflection
%can escape from the scintillator through a side wall 
%and can be registered in the WLS array.
%By placing a reflective surface near the scintillator as shown in~Fig.~\ref{fig:principle}, 
%photons emitted within a cone oriented in the opposite
%direction with respect to the WLS array can be redirected towards the array and can also be registered.
%% FROM FIG.1 CAPTION: 
Scintillation photons which can escape from the scintillator and reach the WLS array are 
emitted within two identical cones - forward and backward one -  with   
the opening angle equal to twice the critical angle in the scintillator material.
A specular surface is used for reflecting photons emitted within the backward cone towards the WLS array.
In our test set-up, for an interaction point located  on the scintillator axis, 
scintillation photons emitted in the forward and backward cone 
illuminate approximately 4 and 10 WLS strips, respectively.
The coordinate of the interaction point along the scintillator bar (axial coordinate) 
can be determined on the basis
of amplitudes measured in individual WLS strips e.g. by the center of gravity method.

The use of WLS strips for the readout of inorganic crystals in PET scanners was proposed in 
\cite{Belcari_NIM2001, Braem_NIM2007}.
However, the applicability of this method in the case of~plastic scintillators is
not obvious due to a substantially smaller light yield 
compared to inorganic crystals which is typically about 10000 photons per MeV energy deposition 
whereas for LYSO crystals it amounts to 32000 photons per MeV.
Moreover, in plastic scintillators, gamma quanta with energy in the order of 1~MeV interact  almost 
exclusively via the Compton effect and hence deposit only a part of their energy.
Therefore, in order to demonstrate the applicability of 
WLS strips for determining the position of the gamma quantum interaction points in a plastic scintillator, we have built a test set-up for measuring the position of interaction of annihilation photons from a $^{22}$Na isotope.
The set-up and the measurement results are described in the next two
sections, respectively, and the paper is closed with conclusions.

\section{Experimental set-up}

\begin{figure}
\centering
\includegraphics[width=0.9\linewidth]{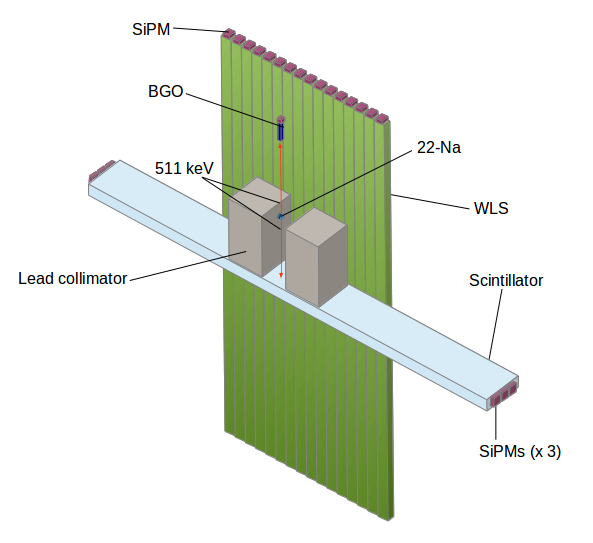}
\caption{Schematic view of the test set-up. 
The WLS strips are arranged orthogonally to the scintillator bar
which is placed at the centre of the length of the WLS strips. A $^{22}$Na source is placed at a distance of 30\,mm above the scintillator. The radioactive part of the source takes the form of a disc with a diameter of about 1\,mm. The BGO detector is placed at a distance of about 60\,mm from the scintillator. A straight line defined by the centre of the BGO crystal and the $^{22}$Na source intersects
the scintillator bar in the center of its width. 
}
\label{fig:setup}
\end{figure}

A bar of BC-420 scintillator with a length of 300\,mm, a width of 19\,mm and a thickness of 5\,mm 
is used for the registration of gamma quanta, and this is shown schematically in Fig.~\ref{fig:setup}.
The type and dimensions of the scintillator bar are identical with the ones used in the J-PET prototype scanner~\cite{Moskal_NIM2014}. 
The scintillator bar was read out on each end by three Hamamatsu S12572-050P silicon photomultipliers (SiPMs).
Photons escaping from the bar through a~5\,mm wide side wall were registered
with an array of 16 WLS strips placed parallel to the wall with a 1.5\,mm air
gap left in between the WLS strips and the scintillator.
The remaining three side walls of the scintillator were covered with reflective aluminium foil.

The WLS strips were 100\,mm long, their width was 5\,mm and the thickness was 3\,mm. 
%(see Fig.~\ref{fig:principle} for definition of the width and thickness of the strips).  
Neighbouring strips were separated by 0.1\,mm gap and thus the active area covered by the strips
along the scintillator bar was 81.5\,mm  (= $16 \times 5$\,mm$+15 \times 0.1$\,mm).

Each WLS strip was read out at one end with a Hamamatsu S12572-050P SiPM having an
active area of $3 \times 3$\,mm$^2$ and thus covering 60\% of the end face of the strip ($3 \times 5$\,mm$^2$).
The opposite end of the strip was covered with  aluminium foil that reflected
re-emitted light towards the SiPM.

The WLS strips were made of BC-482A with three times enhanced dye
concentration.
For increasing the light absorption, an aluminium foil was placed behind
the WLS array in order to reflect back the unabsorbed photons.
%The absorption spectrum of BC-482A covers a range of 350-460\,nm, however, a strong absorption
%is limited to a range of about 390-460\,nm \cite{BC482A}. 
The wavelength of maximum emission of the BC-420 scintillator equals 391\,nm
and thus the emission spectrum only  partly overlaps with the range of strong absorption of BC-482A, 
which is 390-460\,nm~\cite{BC482A}.
Based on measurements of the light transmission coefficient in ELJEN EJ-280 WLS \cite{Braem_NIM2008}, which is 
equivalent  to BC-482A, 
we estimate that about 75\% of the scintillation photons emitted orthogonally to the WLS array are absorbed 
in the array in our test set-up.
%we estimated that in our test set-up about 75\% of scintillation photons emitted
%orthogonally to the WLS array is absorbed in the array.

The wavelength of maximum emission of the BC-482A WLS
equals 494\,nm and fits well in the region of the highest photon detection efficiency 
of 400-500\,nm of the applied SiPMs~\cite{Hamamatsu}.

Test measurements were performed with annihilation photons from a $^{22}$Na source. 
For tagging of the photons we used a small scintillation detector containing a BGO crystal 
of $2.2 \times 2.2$\, mm$^2$ cross section and a length of 10\,mm,
read out by a SiPM.
The 1275\,keV gamma quanta from the $^{22}$Na source were suppressed
by a lead collimator with a 4\,mm slit.

\begin{figure}
\centering
\includegraphics[width=0.8\linewidth]{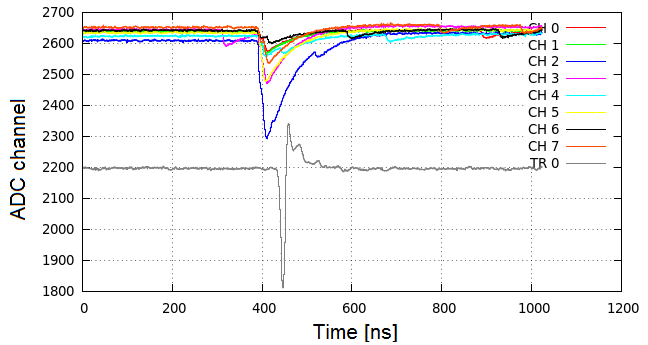}
\caption{Waveforms registered within a 1000\,ns time window 
for eight consecutive WLS strips and for the scintillator bar.
The latter is shown in the lower part of the graph.
In addition to a~group of the WLS pulses 
correlated with the scintillator pulses, there are also a few random pulses with small amplitudes 
resulting from the dark counts of the SiPMs.}
\label{fig:waveform}
\end{figure}

Signals of the SiPMs coupled to the WLS strips were amplified with Advatech-UK AMP-0604 
amplifiers~\cite{Advatech}  
and were sampled with CAEN Digitizer DT5742~\cite{DT5742} with a frequency of 1\,GHz
and a resolution of 12 bits. 
Readout of the digitizer was triggered by a coincidence of pulses from  the plastic
scintillator and the BGO detector, ensuring selection of annihilation
quanta from the $^{22}$Na source.
Typical pulses registered with the digitizer are shown in Fig.~\ref{fig:waveform}.

\section{Results}

\begin{figure}
\centering
\includegraphics[width=0.6\linewidth]{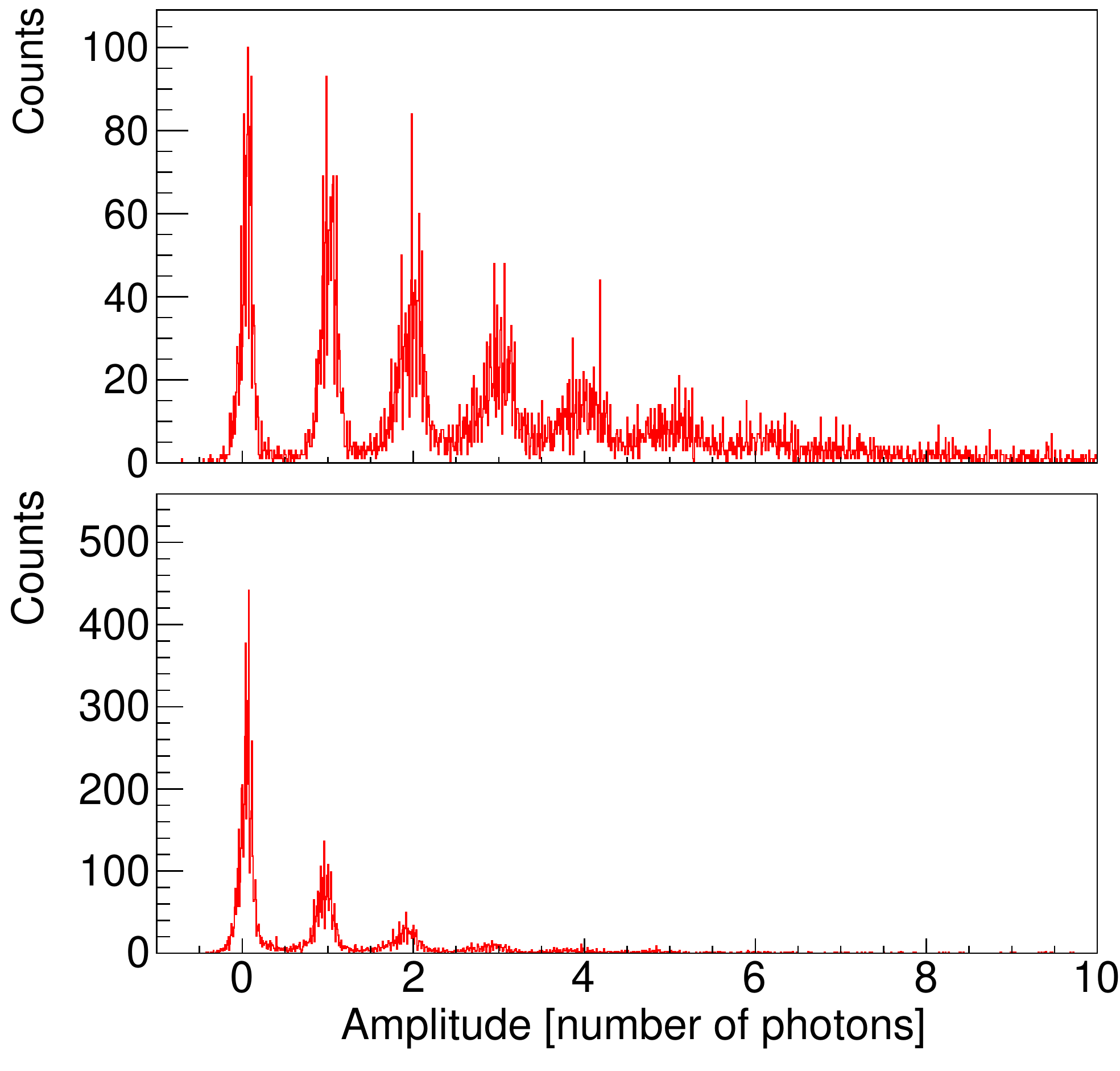}
\caption{Spectrum of amplitudes for a WLS strip lying in front of the interaction point (upper panel)
and for a strip lying aside at a distance of 30\,mm (lower panel).
}
\label{fig:amplitudes}
\end{figure}

In the offline analysis, the recorded samples of the WLS and scintillator pulses 
were used to determine their amplitudes and arrival times.
An example of amplitude spectra for two WLS strips - one located in front of the interaction point
and the other lying 30\,mm aside - is shown in Fig.~\ref{fig:amplitudes}.  
The mean number of photons registered in these strips is 3.8 and 0.8, respectively.

\begin{figure}
\centering
\includegraphics[width=0.6\linewidth]{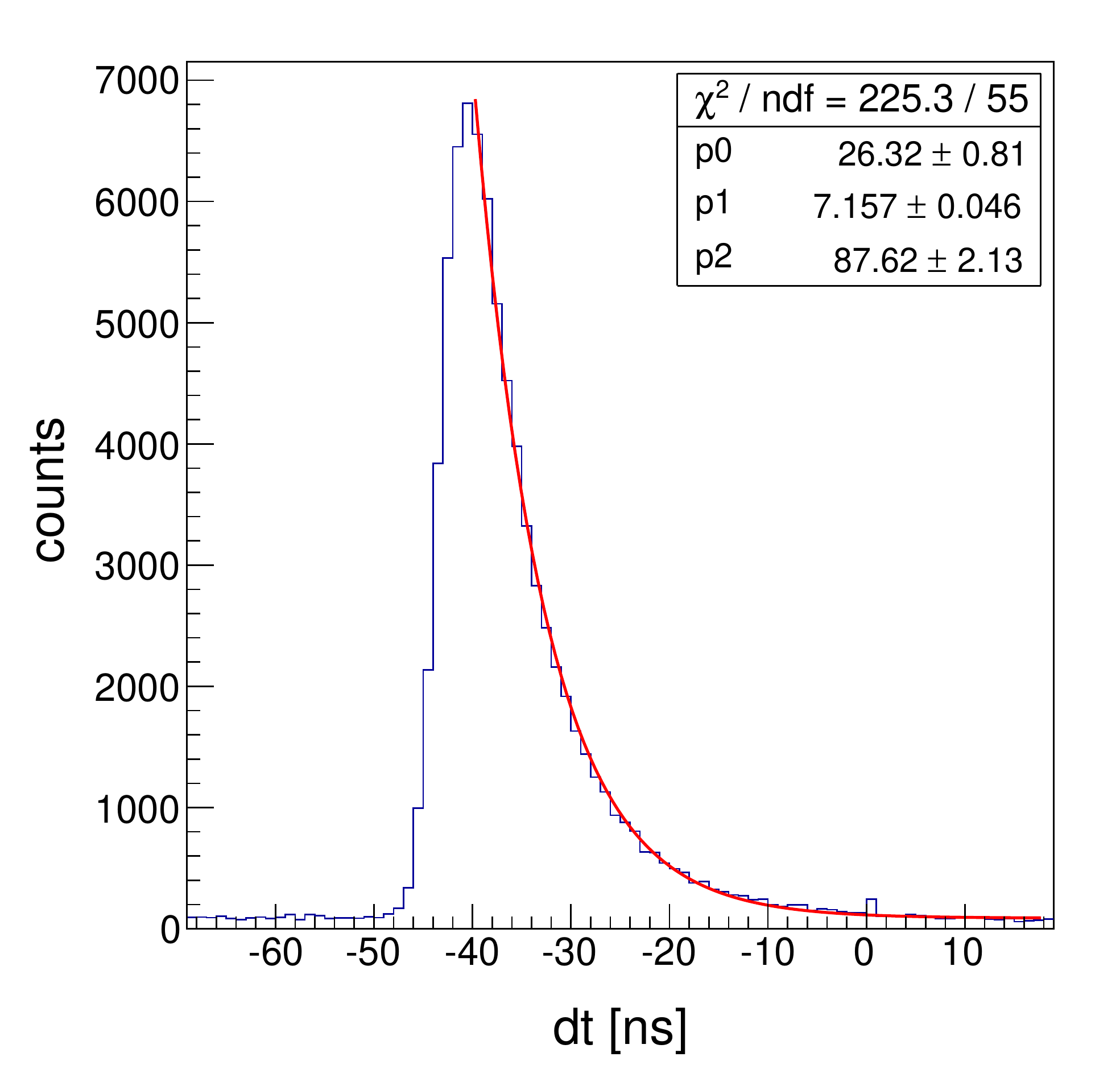}
\caption{Distribution of time difference between the scintillator pulse and the WLS pulses.
Also shown is an exponential curve 
%of the form $y=p0 \cdot exp(-dt/p1) + p2$ 
fitted to the falling edge of the distribution.}
\label{fig:deltat}
\end{figure}

A distribution of time differences between the scintillator pulse and the WLS pulses
is shown in Fig.~\ref{fig:deltat}.
A fit to the falling edge of the distribution with a~function $a \cdot $exp$(-\Delta t/\tau) +b$ gives
a decay time $\tau$ of 7\,ns which is shorter than the 12\,ns decay time of the BC-482A plastic~\cite{BC482A}.
This difference can be explained by the fact that for a few re-emitted photons registered in a~WLS strip, 
only the first one was taken into account in the time measurement based on leading edge discrimination.

In order to reduce accidental coincidences caused by the dark count rate of the SiPMs ($\sim$1\,MHz)
a 50\,ns time window was set on time differences between the scintillator pulse and the WLS pulses.

\begin{figure}
\centering
\includegraphics[width=0.6\linewidth]{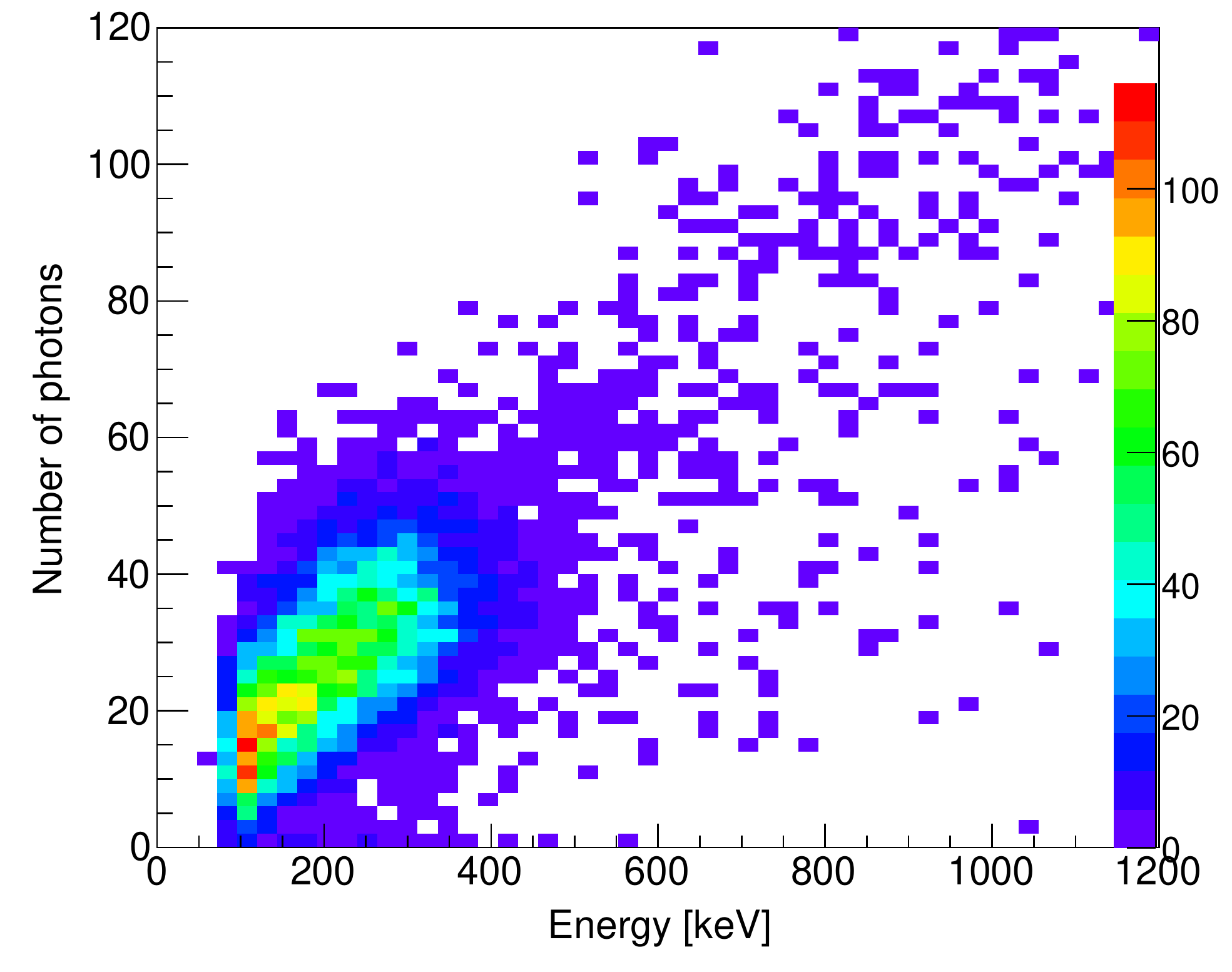}
\caption{Sum of amplitudes (i.e. total number of photons) 
measured in the WLS strips presented as a function of
energy deposited in the scintillator bar. The visible cut out of the lowest amplitudes 
of the scintillator signals 
results from application of a discrimination threshold in the triggering electronics. 
The counts visible at energies above the Compton edge for the annihilation photons (341\,keV) 
originate from the 1275\,keV gamma quanta emitted by the  $^{22}$Na source.}
\label{fig:scyntwls}
\end{figure}

The annihilation photons are registered in plastic scintillator detectors via Compton scattering
and the resulting spectrum of amplitudes of detector pulses has a continuous form. 
%We performed energy callibration of the amplitude spectrum of the scintillator pulses 
%based on a position of the Compton edge for the annihilation photons in the spectrum.
An energy calibration of the scintillator pulse amplitude, based on the position of the Compton edge for the annihilation photons in the spectrum, was performed.
The number of scintillation photons  registered in the WLS strips increases with the energy deposited 
in the plastic scintillator as shown in Fig.~\ref{fig:scyntwls}.
The upper limit for the energy loss of the annihilation photons 
due to Compton scattering equals 341\,keV 
and the corresponding number of photons registered in the WLS strips 
is about 38.
This number has to be corrected for the cross talk in the applied Hamamatsu SiPMs, 
which equals about 25\% for the applied overvoltage of 2.5\,V \cite{Hamamatsu}. 
Thus the true  number of registered photons is about 30.

In order to compare an experimentally determined number of photons at the Compton edge with the expectations based on the properties of the detector set-up
we have performed calculations  taking into account the following 
efficiencies:

\begin{itemize}
\item
0.22 - fraction of full solid angle for photons escaping the scintillator
  and reaching the WLS array,
\item
$0.96 \times 0.96=0.92$ - fraction of photons passing the interfaces scintillator-air and air-WLS, 
\item
0.75 - fraction of photons absorbed in WLS,
\item
0.86 - fluorescence efficiency of WLS,
\item 
0.5 - fraction of confined photons propagating towards one end of a WLS strip,
\item
0.60 - coverage of WLS face with SiPM,
\item
0.62 - SiPM fill factor,
\item
0.35 - photon detection efficiency at 500\,nm.
\end{itemize}
The percentage of registered photons obtained as a product of the above efficiencies equals  0.85\%.
The number of scintillation photons produced by electrons in the BC-420
scintillator is equal to about  10000 per 1\,MeV energy deposit~\cite{Pozzi_NIM2004}, 
%For 0.34\,MeV electrons originating from the backward Compton scattering, number of scintillation
%photons equals 3400 and expected number
which equates to 3400 scintillation photons for the 0.34\,MeV electrons originating from the backward Compton scattering. 
The expected number of photons registered in the WLS strips equals 29 which is close  
to the experimental number of 30 photons.

\begin{figure}
\centering
\includegraphics[width=0.6\linewidth]{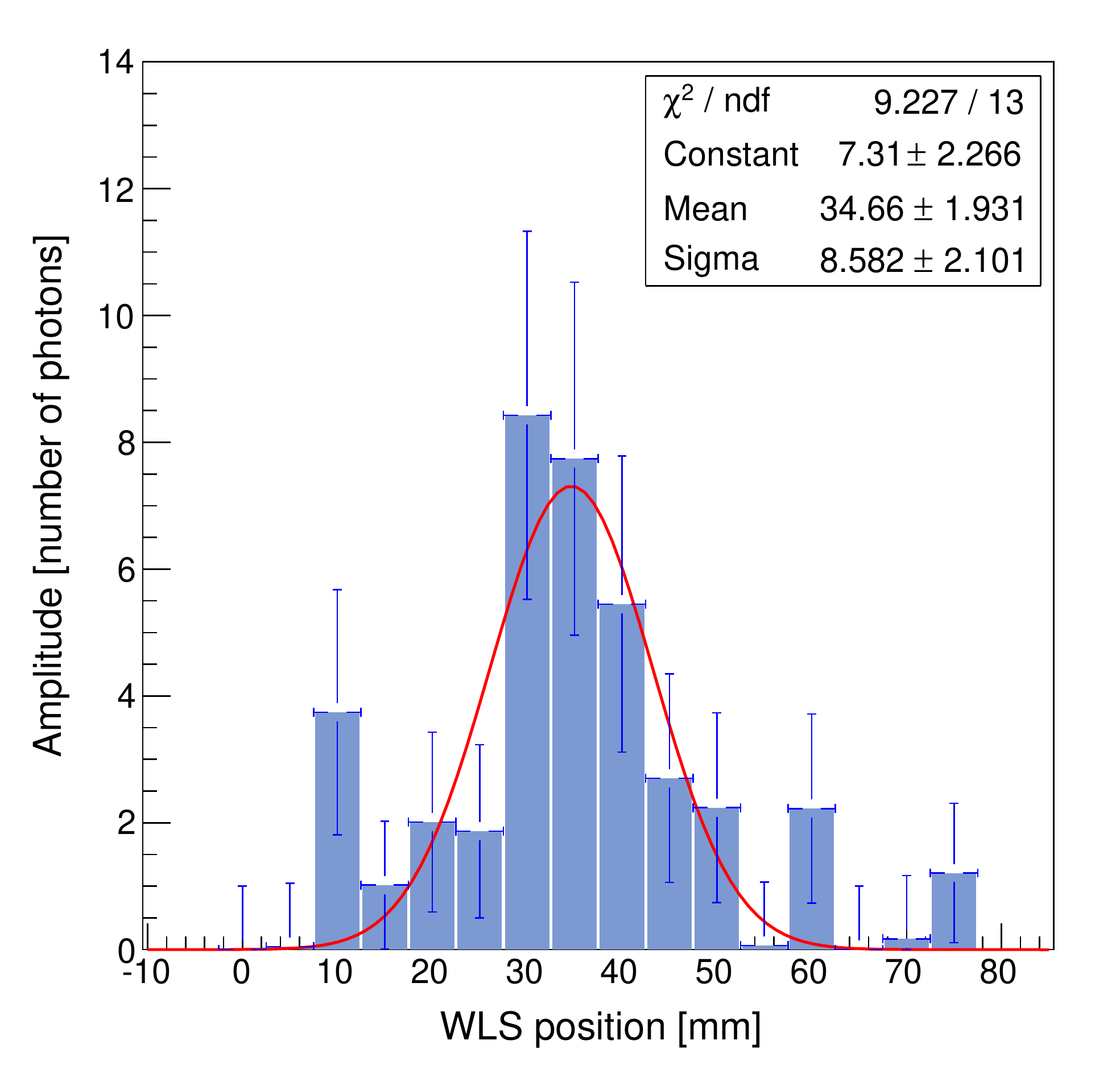}
\caption{Amplitudes measured in individual WLS strips in a typical event
  presented as a function of the strip coordinate 
  and a Gaussian fit applied for determination of the coordinate of the gamma
  quantum interaction point.}
\label{fig:event}
\end{figure}

For each registered event, the distribution of amplitudes of WLS pulses
presented as a function of the WLS coordinates was fitted with a Gaussian function. 
The center of the function was taken as the reconstructed position of the gamma quantum interaction point.
An example of such a fit is shown in Fig.~\ref{fig:event}.

\begin{figure}
\centering
\includegraphics[width=0.6\linewidth]{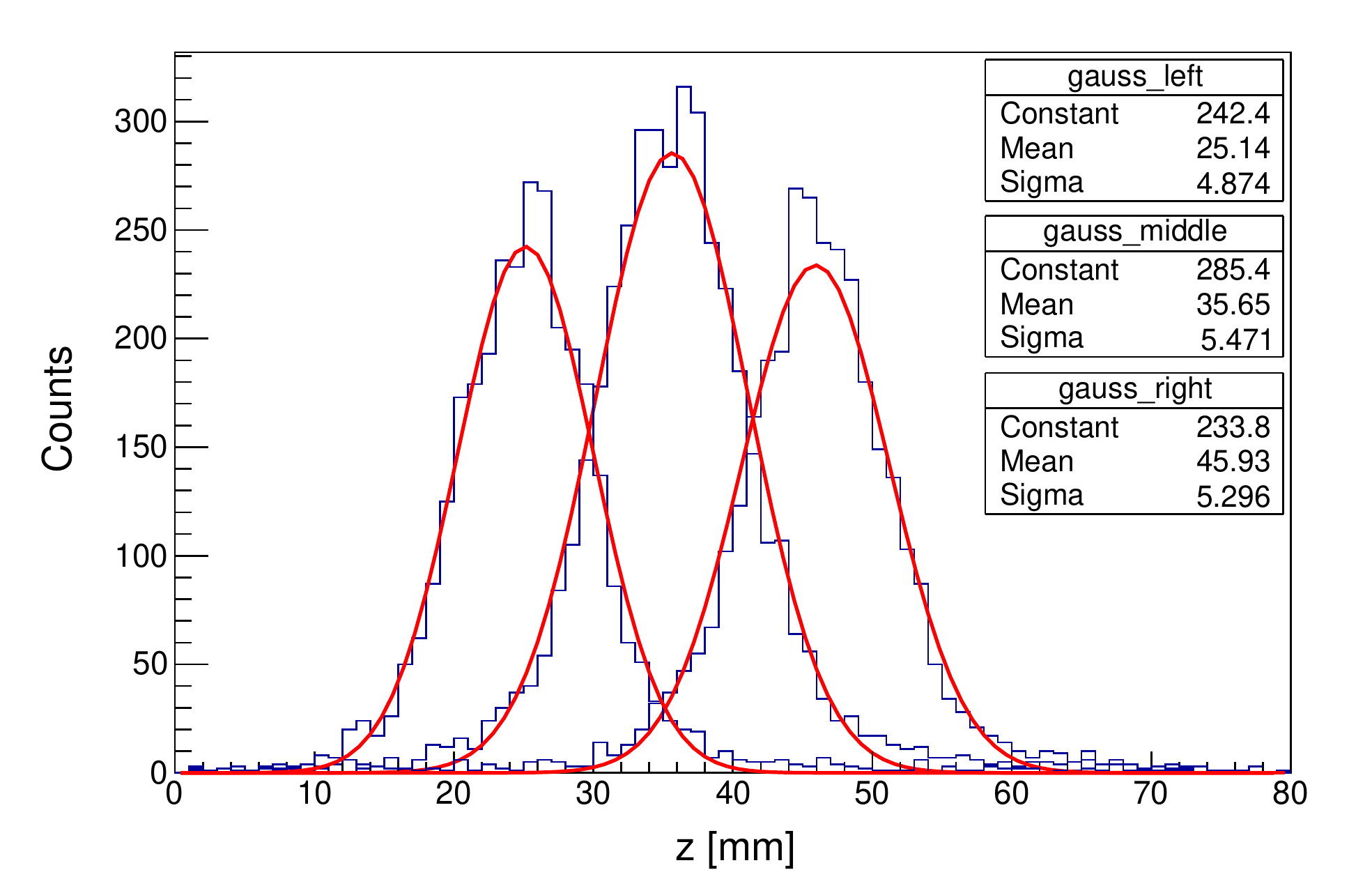}
\caption{Distributions of reconstructed $z$--coordinates of the gamma quantum
  interaction for three different positions of the $^{22}$Na source differing
  by $\Delta z = 10$\,mm. Superimposed lines denote Gaussian fits to the distributions.}
\label{fig:positions}
\end{figure}

Figure~\ref{fig:positions} shows distributions of reconstructed positions for three locations
of the $^{22}$Na source differing by 10\,mm.  
A spatial resolution obtained by fitting the Gaussian function to these distributions is about 5\,mm ($\sigma$).
The centers of the fitted functions reproduce the real positions of the $^{22}$Na source within
a~positioning accuracy of the source of $\pm0.5$\,mm.

For comparison, we also performed a position reconstruction  using
a method based on taking an average coordinate of the WLS strips weighted
with the registered amplitudes of the WLS pulses. 
The average was taken over seven neighbouring strips: a strip with the highest
amplitude and two triplets of~strips located on both sides of it.
The position resolution obtained is of about 6\,mm ($\sigma$) and thus very similar to the one based 
on the Gaussian fit.
 
\section{Conclusions}

We have demonstrated that the interaction point of an annihilation
gamma-quantum in a plastic scintillator bar
can be localized by means of an array of WLS strips measuring
scintillation photons escaping from the scintillator through a side wall.
With our test set-up we reached a position resolution of~5\,mm ($\sigma$) for the
coordinate along the scintillator bar. 
The achieved precision may be further improved e.g. by increasing the coverage of WLS face 
with SiPM (which was only 60\%).
Results presented in this article constitute a basis for construction of detection modules of the J-PET scanner. Each module will consist of a layer of thirteen plastic scintillator bars with dimensions 
of $500 \times 24 \times 6$\,mm$^3$ and a layer of WLS strips. The scintillators will be arranged parallel along the $500 \times 24$\,mm$^2$ sides and the layer of WLS strips will view the $500 \times 6$\,mm$^2$  side of the scintillators. The modules will be used to built a full scale J-PET detector which will enable experimentation in the field of nuclear medical imaging and tests of the discrete symmetries in the decays of positronium atoms 
\cite{Moskal_APP2016, Kaminska_EPJ2016, Gajos_NIM2016}.

\section{Acknowledgements}

We acknowledge the technical support of A.~Mucha and A.~Misiak, and financial support by the Polish Ministry of Science and Higher Education
(Grant No. 2593/7.PR/2012/2).

%\section*{References}

\bibliography{mybibfile}

\end{document}